\documentclass[aip,apl,superscriptaddress,preprint]{revtex4-1}
\usepackage{graphicx}
\usepackage{amsmath}
\usepackage{amssymb}
\usepackage{amsfonts}
\usepackage{filecontents}
\usepackage{color}
\usepackage{soul}
\usepackage[usenames,dvipsnames,svgnames,table]{xcolor}

\begin{document}
\title{Enhanced photocurrent readout for a quantum dot qubit by bias modulation}
\author{J. H. Quilter} \email[Email: ]{j.quilter@sheffield.ac.uk}
\author{R. J. Coles}
\affiliation{Department of Physics and Astronomy, University of Sheffield, Sheffield, S3 7RH, United Kingdom}
\author{A. J. Ramsay}
\affiliation{Department of Physics and Astronomy, University of Sheffield, Sheffield, S3 7RH, United Kingdom}
\affiliation{Hitachi Cambridge Laboratory, Cavendish Laboratory,University of Cambridge, Cambridge CB3 0HE, United Kingdom}
\author{A. M. Fox}
\author{M. S. Skolnick}
\affiliation{Department of Physics and Astronomy, University of Sheffield, Sheffield, S3 7RH, United Kingdom}
\date{\today}
\begin{abstract}
We demonstrate coherent control of a quantum dot exciton using photocurrent detection with a sinusoidal reverse bias. Optical control is performed at low bias, where tunneling-limited coherence times are long. Following this step, the tunneling rates are increased to remove the long-lived hole, achieving a high photocurrent signal. For a detection efficiency of 68\%, electron and hole tunneling times during optical control of 200~ps and 20~ns can be achieved, compared to 120~ps and 7~ns for the constant bias case, respectively.
\end{abstract}
\maketitle

Excitons and single carrier spins isolated in semiconductor quantum dots (QDs) are promising candidates for solid-state qubits as they offer the combination of long coherence times\cite{Kroutvar2004,Greilich2006a}  with the possibility of ultrafast optical control\cite{Press2008}. Photocurrent (PC) detection is an important measurement technique for coherent control experiments, as it allows direct electrical measurement of the occupancy of a single QD. Photocurrent detection can be used to measure both exciton \cite{Zrenner2002,Stufler2006,Muller2011,Takagi2008,Wu2011} and hole-spin states \cite{Godden2012,Muller2012}. It is compatible with optical cavities\cite{Kistner2009}, and  recently, QD avalanche photodiode structures have been demonstrated\cite{Bulgarini2012}.

\begin{figure}
\includegraphics[width=8.48cm]{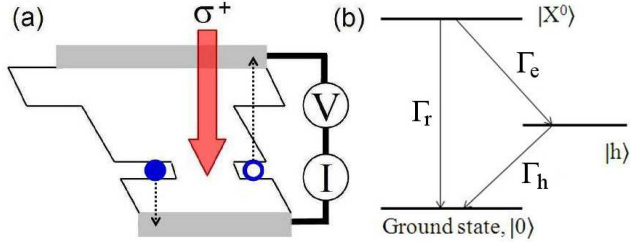}
\caption {Color online. (a) Band structure of the diode under reverse bias. The absorption of a circularly polarized pulse generates an electron (solid circle) and hole (hollow circle), which tunnel through their respective potential barriers. (b) The two decay paths of the neutral exciton $|X^0\rangle$ to the crystal ground state $|0\rangle$, either by radiative recombination or sequential carrier tunneling via the hole state $|\mathrm{h}\rangle$}.
\label{sample}
\end{figure}

To illustrate the principle of photocurrent detection, fig.~\ref{sample}~(a) depicts the resonant excitation of a single QD in a Schottky diode with a laser pulse of area $\pi$. A single electron-hole pair is generated in the QD.  The states of the electron-hole system are shown in fig.~\ref{sample}~b). Since hole tunneling is slow, the coherent lifetime of the neutral exciton $|X^0\rangle$ is dependent on two dominant decay paths to the crystal ground state, $|0\rangle$: radiative recombination; or electron, followed by hole tunneling, resulting in a photocurrent.

Ideally, the contribution to the photocurrent is  one electron-hole pair per pulse\cite{Zrenner2002}, leading to the proposed use of such a device as an optically-triggered single electron source. In practice, there are three factors that limit the photocurrent detection efficiency. Firstly, there is competition between radiative recombination and electron tunneling rates\cite{Fry2000}. Secondly, intensity damping reduces the population inversion\cite{Ramsay2010a}. Thirdly, the heavy-hole tunneling rate can be slower than the repetition frequency of the laser\cite{Kolodka2007,Mar2011} ($T_\mathrm{rep}^{-1}$), leading to Pauli blocking when the next pulse arrives. For an optimized repetition rate, the maximum photocurrent is limited to $\sim e\Gamma_\mathrm{h}$, where $\Gamma_\mathrm{h}$ is the hole tunneling rate. Hence there is a trade-off between the photocurrent signal and the tunneling limited coherence time of the hole.

In this letter, we propose and demonstrate the use of voltage modulation to overcome the trade-off between photocurrent detection efficiency and the carrier tunneling times that limit the coherence of both the exciton and hole spin. To illustrate the principle, consider the scenario of a square-wave modulation applied to the QD photodiode, which switches between reverse voltages $V_\mathrm{low}$ and $V_\mathrm{high}$. During optical manipulation, the relatively slow tunneling rates $\Gamma_\mathrm{e,h}(V_\mathrm{low})$ enable high-fidelity optical control. Afterwards, the tunneling rates are increased to $\Gamma_\mathrm{e,h}(V_\mathrm{high})$, sweeping the carriers from the QD ready for the next laser pulse sequence. The hole tunneling rate $\Gamma_\mathrm{h}(V_\mathrm{low})$ acting on the QD during optical control, and the maximum photocurrent signal $\sim e\Gamma_\mathrm{h}(V_\mathrm{high})$ are decoupled, overcoming the trade-off. To demonstrate the principle, a simpler experiment using a cosine modulation is presented. The reverse bias applied to the QD is described by:
\begin{equation}
V(t,\phi) = V_\mathrm{DC}+V_\mathrm{AC} \cos(2 \pi t/ T_\mathrm{rep}^{-1} -\phi), \nonumber
\end{equation}
where $t$ is the time after the arrival of the laser pulse,  $\phi$ is the phase  and $V_\mathrm{AC}$ is the amplitude of the modulation.

The device consists of a layer of InGaAs QDs  embedded in the intrinsic region of an n-i-Schottky diode structure, with an additional 75~nm AlGaAs barrier to block hole tunneling. A single QD is accessed through a 500~nm diameter aperture in an opaque Al mask. Full details of the sample can be found in ref. \onlinecite{Kolodka2007}. The sample is held at 4.2~K in a helium bath cryostat and is excited at normal incidence with circularly polarized, 0.2~meV FWHM Gaussian pulses, derived by optically pulse-shaping the output from a single 100~fs Ti:sapphire laser with repetition rate $T_\mathrm{rep}^{-1}$ = 76.2~MHz.

\begin{figure}
\includegraphics[width=8.48cm]{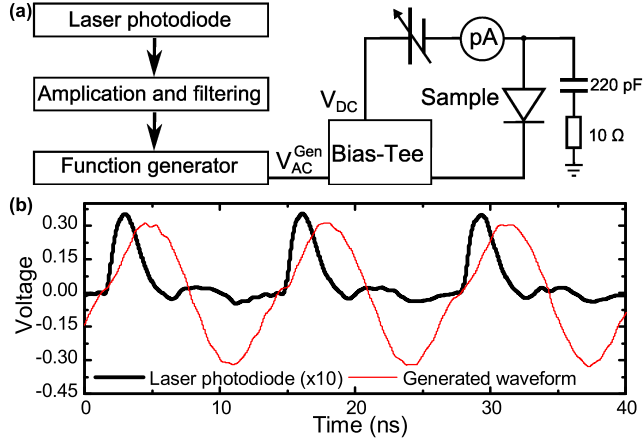}
\caption {(Color online) (a) Circuit for generating an AC-voltage synchronized with the laser pulse. A bias-tee is used to combine AC and DC voltages. (b) Oscilloscope traces showing the laser photodiode output and the resulting triggered AC waveform with ($V_\mathrm{AC}^\mathrm{Gen}$) = 0.30~V.}
\label{setup}
\end{figure}

The system for generating an AC voltage synchronized with the laser pulses is presented in fig.~\ref{setup}~(a), a similar circuit is used by de Vasconcellos {\it et al}\cite{Vasconcellos2010}. The laser output is monitored with a photodiode. This signal is spectrally filtered to remove higher harmonics and amplified to be strong enough to trigger a function generator. Before connecting to the sample, the AC voltage is offset by a DC voltage ($V_\mathrm{DC}$) using a broadband bias-tee. Figure~\ref{setup}~(b) shows typical oscilloscope traces of the laser photodiode output used as an input for the filter circuit, and the resulting synchronized AC-voltage applied to the device.

\begin{figure}
\includegraphics[width=8.48cm]{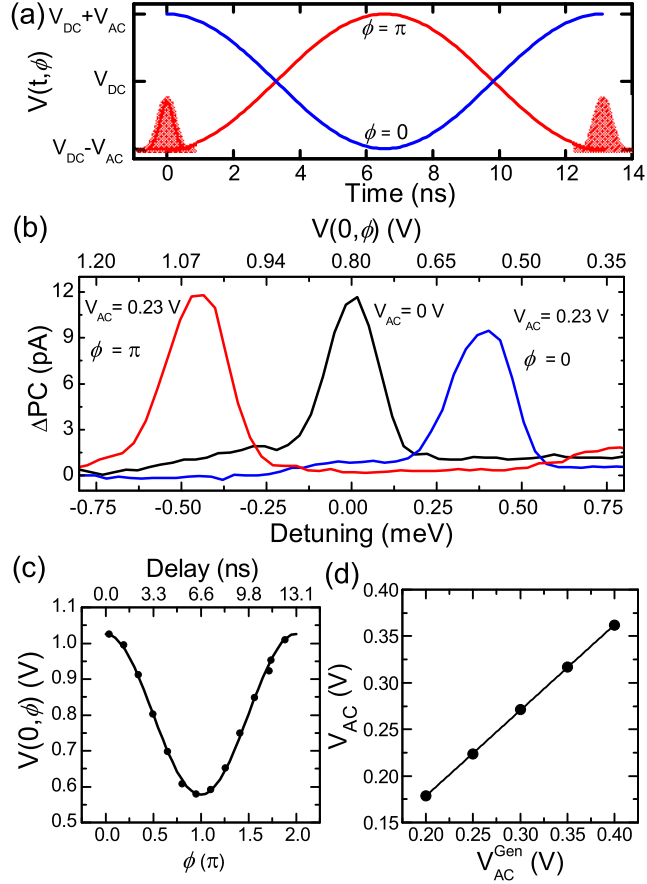}
\caption {(Color online) (a) Time-varying bias, $V(t,\phi)$ with respect to laser pulse arrival, as indicated by the red shaded pulses. (b) Neutral exciton peak in photocurrent spectra with $V_\mathrm{DC}=0.8$~V and $V_\mathrm{AC} = 0.23$~V for $\phi = \pi$ and $\phi =$~0, compared with reference spectrum with $V \equiv V_\mathrm{DC} = 0.8$~V and $V_\mathrm{AC}=0$~V. Background photocurrent is subtracted for clarity. (c) The reverse bias on pulse arrival $V(0,\phi)$ is plotted as a function of $\phi$. (d) AC voltage amplitude ($V_\mathrm{AC}$) as a function of voltage output by the function generator ($V_\mathrm{AC}^\mathrm{Gen}$), yielding a gradient of 0.918~$\pm$~0.004.}
\label{sine}
\end{figure}

Figure 3 (a) presents timing diagrams of the laser pulse with respect to the AC-voltage in the cases where $\phi$~=~0 and $\pi$.  To confirm that an AC voltage is applied to the QD, the energy shift of the neutral exciton peak is measured. For reference, a spectrum with only a DC bias of  $V_\mathrm{DC} = 0.8$~V, ({\it i.e.} $V_\mathrm{AC} = 0$~V) is  presented in fig.~\ref{sine}~(b). By applying an additional AC-voltage of $V_\mathrm{AC}= 0.23$~V, the photocurrent peak is shifted by the Stark effect \cite{Fry2000a} according to the phase, with the extreme energy shifts achieved for $\phi=0, \pi$ as presented in fig.~\ref{sine}~(b). When $\phi = \pi$, $V(0, \pi)$ is at its minimum value, shifting the peak to lower energy compared to the DC case, with $V_\mathrm{AC} = 0$. The opposite is true for $\phi = 0$ and the peak is shifted to higher energy. The voltage ($V(0,\phi)$) applied to the QD at the time of arrival of the laser pulse is deduced from the energy of the photocurrent peak using the quantum-confined Stark-shift measured for DC voltages\cite{Alen2003a}. As shown in fig.~\ref{sine}~(c), the voltage applied on arrival of the laser pulse, $V(0,\phi)$ oscillates as a function of phase $\phi$ with a frequency of $76.6\pm 0.3$~MHz, confirming that the AC-voltage is effectively applied to the QD.

To measure the transfer of the applied AC-voltage to the QD, $V_\mathrm{AC}$ is extracted from the fit presented in fig.~\ref{sine}~(c) and plotted as a function of the amplitude of the AC voltage specified by the function generator, $V_\mathrm{AC}^\mathrm{Gen}$, in fig.~\ref{sine}~(d). From the gradient we deduce that 92\% of the applied voltage modulation amplitude is transferred to the QD, suggesting that impedance matching is excellent. This is consistent with the estimated 0.9~ns RC-time of our unoptimized device.

\begin{figure}
\includegraphics[width=8.48cm]{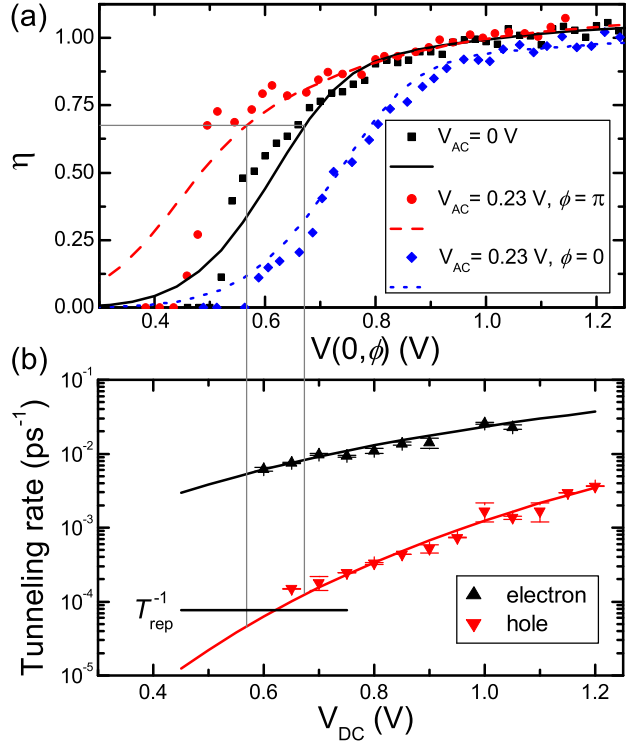}
\caption {(Color online)(a) Detection efficiency ($\eta$) of the neutral exciton peak in photocurrent spectra as a function of instantaneous voltage at pulse arrival, $V(0,\phi)$. $\eta$ for a series of voltage-dependent spectra with DC-only bias are presented for comparison. The photocurrent amplitude is enhanced by the AC modulation for the same $V(0,\phi)$, with $V_\mathrm{AC}$ = 0.23~V and $\phi =\pi$. (b) Electron and hole tunneling rates as a function of voltage. Grey lines match the points where $\eta = 0.68$ in (a) to the corresponding instantaneous tunneling rates in (b).}
\label{amplitude}
\end{figure}

Figure~\ref{amplitude}~(a) compares the detection efficiencies achieved using a DC-only bias, and the AC-modulation scheme with $V_{AC}$ = 0.23~V and $\phi=0,\pi$. The amplitude of the photocurrent peak generated by a $\pi$-pulse, $PC(\pi)$ is measured as a function of effective voltage on pulse arrival, $V(0,\phi)$. The detection efficiency is defined as $\eta = PC(\pi)/(T_\mathrm{rep}^{-1} \times e)$. A photocurrent of 12.2~pA corresponds to one electron per pulse, {\it i.e.} $\eta = 1$. In the case of a DC bias,  $V_\mathrm{AC}=0$, the detection efficiency only reaches unity when $ V_\mathrm{DC}$ is greater than 1.0~V. For $V_{\mathrm{DC}}$ below this value, due to slow hole tunneling, there is a high probability that the dot is occupied by a hole on arrival of the next pulse, preventing  absorption of the laser pulse due to Coulomb-shift in the absorption energy. A sharp threshold is observed at $V_\mathrm{DC} \approx$ 0.52~V, below which the photocurrent amplitude is negligible. The threshold is attributed to charging of the dot by a single electron, suppressing the neutral exciton transition\cite{Warburton1997}.

In fig.~\ref{amplitude}~(a) the x-axis is $V(0,\phi)$, the voltage applied on arrival of the laser pulse. This enables a direct comparison of the detection efficiencies achieved for particular values of the electron and hole tunneling rates experienced by the QD during optical manipulation.
The detection efficiency is enhanced when $V_\mathrm{AC} \neq 0$, $\phi = \pi$  for $V(0,\pi)$ \textless~0.8~V, increasing the bias range where a clear signal can be measured.
Here the laser pulse arrives at the minimum in the applied voltage. Afterwards, the hole tunneling rate is increased via the increasing reverse voltage. The hole is swept from the QD ready for the next laser pulse to be absorbed increasing the detection efficiency. The photocurrent is suppressed for all bias values below 1.2~V when $\phi$ =~0. Here the laser pulse arrives at the maximum in the applied reverse voltage. Afterwards, the hole tunneling rates are decreased, increasing the probability that the hole remains in the dot when the next laser pulse arrives, thereby blocking absorption, and hence reducing the photocurrent.

In order to quantify the improvement in tunneling rates for a given detection efficiency, fig.~\ref{amplitude}~(b) presents the electron ($\blacktriangle$) and hole (\textcolor{red}{$\blacktriangledown$)} tunneling rates measured for a DC-only bias using an inversion recovery technique as in ref. \onlinecite{Kolodka2007}.  A pair of $\pi$-pulses excite the neutral exciton and the photocurrent as a function of inter-pulse time-delay is measured. The tunneling times are extracted from the biexponential saturation of the sum of the photocurrent signals measured for co and cross-circular excitation. The lines are fits to a WKB-expression for the tunneling rates: $\Gamma_\mathrm{e(h)}(V) = \Gamma_\mathrm{e0(h0)}\exp(-V_\mathrm{e0(h0)}/(V+V_\mathrm{bi}))$, from ref.~\onlinecite{Fry2000a}, where $V$ is the reverse bias, $V_\mathrm{bi}=0.76~\mathrm{V}$ is the built-in voltage and $\Gamma_\mathrm{e0(h0)}$, $V_\mathrm{e0(h0)}$ are constants.

To test our understanding of the AC-modulation scheme, the detection efficiency is calculated and presented as the lines in fig.~\ref{amplitude}~(a). The rate equation model is illustrated in fig.~1~(b) and considers the population of the crystal ground, exciton and hole states\cite{Kolodka2007}. The carrier tunneling rates $\Gamma_\mathrm{e,h}(V)$ depend on the instantaneous applied voltage $V(t,\phi)$, and the radiative recombination is  assumed to be independent of voltage due to the relatively small shift in energy over the voltage range used here\cite{Fry2000b}. The calculated photocurrent detection efficiency is given by $\eta = f(\Gamma_\mathrm{h}) \times g(\Gamma_\mathrm{e})$, where:
\begin{eqnarray}
f(\Gamma_\mathrm{h}) &\approx& 1-\exp\bigg(-\int_{0}^{T_\mathrm{rep}}\Gamma_\mathrm{h}(t)\,\mathrm{d}t\bigg)\nonumber\\
\text{and}\nonumber\\
g(\Gamma_\mathrm{e}) &=& \int_{0}^{T_\mathrm{rep}} \Gamma_\mathrm{e}(t) \exp\bigg( -\int_0^{t}(\Gamma_\mathrm{r}+\Gamma_\mathrm{e}(\tau)){\, \mathrm{d}\tau}\bigg)\,\mathrm{d}t.\nonumber
\end{eqnarray}

Here, $f(\Gamma_{\mathrm{h}})$ is the probability that the hole no longer remains in the QD when the next pulse arrives and $g(\Gamma_{\mathrm{e}})$ is the probability that the exciton decays via electron tunneling instead of radiative recombination. An inverse radiative recombination lifetime of $\Gamma_\mathrm{r}$ = 1/(400~ps) was extracted from a fit to the DC-only bias data. This value is consistent with literature values for the radiative recombination of InAs/GaAs QDs\cite{Langbein2004}.

The detection efficiency predicted by the model closely matches the experimental data for higher voltages $(V(0,\phi)$~\textgreater~0.6~V). However, the model does not explain the threshold  behavior observed in the photocurrent at lower bias, which, as explained above, is related to a single electron charging threshold.

As an illustration of the improvement in the coherence times that can be achieved by using the AC-modulation scheme, we note that to achieve a detection efficiency of $\eta=0.68$ (as indicated by the horizontal grey line in fig.~\ref{amplitude}~(a)), a DC bias of 0.67~V is required. However, by applying an AC-modulation ($V_{AC}$ = 0.23~V  $\phi=\pi$) $V(0,\pi)$ is reduced by 0.1~V to 0.57~V. The grey vertical lines in fig.~\ref{amplitude} link the points in fig.~\ref{amplitude}~(a) where $V(0,\phi) = 0.67, 0.57$~V to the corresponding electron and hole tunneling rates in fig.~\ref{amplitude}~(b) for equivalent DC biases. The reduced voltage when the pulse is absorbed means that during coherent optical control, the carrier tunneling times are  $\Gamma_\mathrm{e}^{-1}=200$~ps and $\Gamma_h^{-1}=20$~ns, compared to 120~ps and 7~ns for the DC-only bias. In principle, the effective coherence times of the exciton and the hole spin have been approximately doubled at no cost to the detection efficiency. Alternatively, a detection efficiency of 50\% can be achieved for an effective hole tunneling rate of 100~ns, which would be impossible for a DC-only bias.

\begin{figure}
\includegraphics[width=8.48cm]{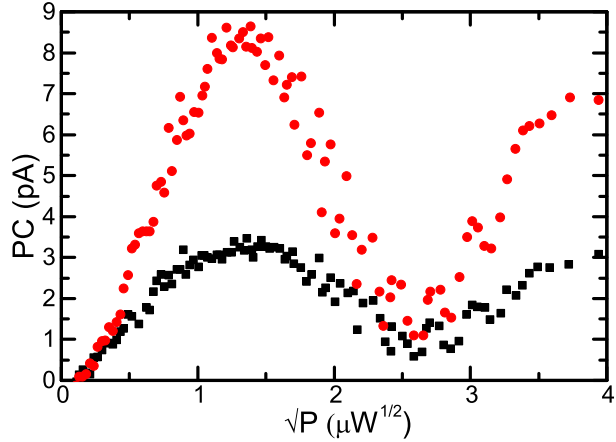}
\caption {(Colour online.) Rabi rotations of the neutral exciton transition, for a DC bias of 0.530~V ($\blacksquare$) and with AC modulation ($V_\mathrm{DC}$ = 0.842~V, $\phi = \pi$ and $V_\mathrm{AC}$ = 0.312~V) (\textbf{{\color{red} \textbullet}}). P is the laser power transmitted to the sample. A linear background photocurrent with laser power is subtracted from the data.}
\label{rabi}
\end{figure}

To test that the AC-modulation does not degrade the contrast of an exciton Rabi rotation, Rabi rotations  of the neutral exciton transition for DC-only bias and for AC modulation are compared in fig.~\ref{rabi}. For both traces, $V(0,\phi)$ = 0.530~V, as this provides the largest increase in detection efficiency $\eta$, compared with the DC-only bias situation as can be seen in fig.~\ref{amplitude}~(a). The amplitude of the AC-modulated Rabi-rotation is  increased to 8.4~$\pm$~0.3~pA compared to the DC case 3.3~$\pm$~0.2~pA for a pulse-area of $\pi$.  Additionally, the contrast of the rotations defined as $(PC(\pi)-PC(2\pi))/PC(\pi)$ for the case with AC modulation is 0.83~$\pm$~0.06 and 0.70~$\pm$~0.1 for the DC-only bias. As the two contrasts are similar within experimental error, we infer that on the timescale of the laser pulse, no additional dephasing is introduced by the AC voltage.

In summary, we demonstrate a voltage modulation scheme for overcoming the compromise between carrier tunneling and detection efficiency encountered in coherent control experiments using the photocurrent detection technique. In this scheme, coherent control is performed at low reverse bias where exciton coherence times are long. Afterwards, the bias is increased to sweep the carriers from the dot to be detected as a photocurrent. Future directions include experiments with more complex voltage waveforms, where electrical switching of the tunneling rates could be used to optimize hole spin initialization, control and detection using the photocurrent technique 
\newline

The authors acknowledge the EPSRC (UK) EP/G001642 and EP/J007544/1 for financial support, and H. Y. Liu and M. Hopkinson for sample growth.
\bibliography{libraryAC}
\end{document}